\documentclass{elsart}

\usepackage{natbib}

\usepackage{graphicx}
\usepackage{amssymb,amsmath} 
\usepackage{graphicx}
\usepackage{ulem}
\usepackage{units} 

\usepackage{units} 

\providecommand{\bc}{\begin{center}}
\providecommand{\ec}{\end{center}}
\providecommand{\be}{\begin{equation}}
\providecommand{\ee}{\end{equation}}
\providecommand{\bea}{\begin{eqnarray}}
\providecommand{\eea}{\end{eqnarray}}
\providecommand{\bdm}{\begin{displaymath}}
\providecommand{\edm}{\end{displaymath}}
\providecommand{\bdma}{\begin{eqnarray*}}
\providecommand{\edma}{\end{eqnarray*}}
\providecommand{\ba}{\begin{eqnarray*}}
\providecommand{\ea}{\end{eqnarray*}}
\providecommand{\bi}{\begin{itemize}}
\providecommand{\ei}{\end{itemize}}
\providecommand{\benum}{\begin{enumerate}}
\providecommand{\eenum}{\end{enumerate}}

\providecommand{\refkl}[1]{(\ref{#1})}

\providecommand{\twoCases}[4]{
  \left\{ 
    \begin{array}{ll} 
      #1 & #2 \\
      #3 & #4 
    \end{array} 
  \right.
}

\providecommand{\myVector}[1]{
  \left(\begin{array}{c} 
      #1 
   \end{array} \right)
}

\providecommand{\text}[1]{{\mbox{ #1}}}

\providecommand{\tr}{^{\text{T}}}

\providecommand{\fig}[2]{
   \begin{center}
     \includegraphics[width=#1]{#2}
   \end{center}
}

\providecommand{\abl}[2]{\frac{{\rm d} #1}{{\rm d} #2}}  

\providecommand{\sub}[1]{_{\rm #1}}
\renewcommand{\sup}[1]{^{\rm #1}}

\begin{document}
\begin{frontmatter}
\title{Comparing Numerical Integration Schemes for Time-Continuous
  Car-Following Models}
\author[TUD]{Martin Treiber\corauthref{cor1}}, \ead[url]{http://www.mtreiber.de} \ead{treiber@vwi.tu-dresden.de} 
\author[TIT]{and Venkatesan Kanagaraj}

\address[TUD]{Technische Universit\"at Dresden, Institute for
  Transport \& Economics,\\ W\"urzburger Str. 35, D-01187 Dresden,
  Germany}
\address[TIT]{Technion–Israel Institute of Technology, Haifa 32000, Israel.}
\corauth[cor1]{Corresponding author. Tel.: +49 351 463 36794; fax: +49 351 463 36809}

\begin{abstract}
When simulating trajectories by integrating time-continuous car-following models,
standard integration schemes such as the forth-order Runge-Kutta
method (RK4) are rarely used while the simple  Euler's
method is popular among researchers. We compare four explicit methods:
Euler's method, ballistic update, Heun's method (trapezoidal rule), and
the standard forth-order RK4. As performance
metrics, we plot the global discretization error as a
function of the numerical complexity. We tested the methods on several time-continuous 
 car-following models in several multi-vehicle simulation
 scenarios with and without discontinuities such as stops or a
 discontinuous behavior of an external leader. 
We find that the theoretical advantage of RK4 (consistency order~4) only plays
 a role if both the acceleration function of the model and the
 external data of the simulation scenario are sufficiently often
 differentiable. Otherwise, we obtain lower (and often fractional)
 consistency orders. Although, to our knowledge,  Heun's method has never been used for
 integrating car-following models, it turns out to be the
 best scheme for many practical situations. The
 ballistic update always prevails Euler's method although
 both are of first order.
\end{abstract}

\end{frontmatter}

\section{Introduction}
Time-continuous car-following models (or more precisely, their longitudinal dynamics
components)  prescribe the acceleration of
individual cars as a function of the driver's characteristic behaviour
and the
surrounding traffic. Formally, their mathematical formulation is
equivalent to that of physical particles following Newtonian dynamics
with the physical forces replaced by ``social forces''~\citep{HelbTilch-98ab}. 
In contrast to car-following models formulated in discrete
time (coupled maps) or fully discretely (cellular
automata), time-continuous car-following models must be
augmented with a numerical integration method in all
but the most trivial analytically solvable cases~\citep{TreiberKesting-Book}.

Mathematically speaking, time-continuous car-following models without
explicit reaction time delay represent coupled ordinary
differential equations (ODE).
The most popular models of this class are the optimal-velocity
model (OVM) of~\cite{Bando1}, derivatives such as the full-velocity
difference model (FVDM) of~\cite{Jiang-vDiff01}, and the
Intelligent-Driver Model (IDM) by~\cite{Opus}. Early car-following
models such as that by Gazis, Herman and Rothery (GHR,\cite{gazis1961nonlinear})
or the linear ACC model by~\cite{Helly} also fall into this class.

Because of the complexity and possible event-oriented components of
traffic flow scenarios, one generally assumes a fixed common time step
$h$ and explicit numerical
schemes to obtain trajectories from the model equations. 
 In the general literature on numerical mathematics (see, e.g.,~\cite{NumMathe-Springer}),
the standard explicit numerical integration scheme for ODEs is the 
fourth-order Runge-Kutta method (RK4). However, in 
the domain of microscopic traffic flow modelling, the use of this
method is rarely stated (counterexamples include~\cite{Kaupuzs-2004}
and~\cite{ShamotoPRE-2011}). Instead, most authors  apply
simpler methods or do not specify the numerical method at all. 
Commonly used schemes are the simple Euler
method~\citep{Aw-Rascle-2002} or the ballistic update assuming constant
accelerations during one time
step~\citep{TreiberKesting-Book}. Notice that also the open-source traffic simulators
SUMO~\citep{SUMO2011} and AIMSUN~\citep{AIMSUN2010} use simple Euler update for the
positions. 

Sometimes, time-continuous
models plus a lower-order update method are proposed as 
time-discrete models in their own right. For example, Newell's microscopic
model~\citep{Newell} corresponds to an OVM with a triangular
fundamental diagram and simple 
Euler-update with a time step $h$ equal to
the desired time gap~\citep{TreiberKesting-Book}.  Similarly, a 
time-discrete model has been 
derived from the GHR model by using Euler update
with $h$ equal to the 
reaction time, and qualitatively different behavior has been found
compared to integrating the original GHR model with the RK4
method~\citep{Jamison-2009}. In the context of car-following methods,
the ballistic method is particularly appealing since it allows
to model reaction times 
without introducing explicit delays which would transform the
ODEs of time-continuous car-following models (such as all time-continuous models mentioned
above) into
delay-differential equations. It has been
shown~\citep{ThreeTimes-07} that integration of the IDM by the ballistic method
with time step $h$ is essentially equivalent to an explicit reaction
time delay $T_r=h/2$ of the corresponding delay-differential equations 
(which are, then, integrated
by higher-order methods or very small time steps).

Nevertheless, it is often desired to approach the true solution of
time-continuous car-following models as closely as possible.
A criterion for the quality of an integration scheme is its (local or
global) consistency order stating how fast the approximate numerical
solution converges to the true solution when decreasing the time step
$h$ (\cite{NumMathe-Springer}, see Sect.~\ref{sec:math} for details).
However, for practical
integration steps $h$, higher-order methods do
not necessarily lead to lower discretization errors. Moreover, if the
acceleration function of the model
is not sufficiently smooth (differentiable) or the
simulation scenario contains discontinuities such as stops, lane
changes, or traffic lights, the actual consistency
order of a given numerical scheme is generally lower than its nominal
order~\citep{NumMathe-Springer}. Finally, higher-order methods need
several evaluations of the 
model's acceleration function per vehicle and per time step while
Euler's method and the 
ballistic scheme need only one. 

This leads to
following question: 
``Does the higher numerical accuracy of higher-order schemes
  outweigh their higher numerical complexity in terms of computation time, for
  practical cases?''
Specifically, we would like to know which numerical scheme has the
lowest global discretization error for a given numerical complexity,
and how this 
depends on the model and the simulation scenario.

In this work, we profile four numerical methods, simple Euler, ballistic
scheme, Heun's rule or trapezoidal rule, and RK4, for three car-following models
(OVM, FVDM, IDM) in several multi-vehicle simulation scenarios. We
found that RK4 is, in fact, superior if
certain rather restrictive conditions for the differentiability of the
acceleration function and the  external data are satisfied, and if a
high numerical precision is required. In most
practical situations, however, the 
ballistic scheme and the trapezoidal rule turn out to be
the most efficient and robust methods, although the latter is rarely
used.  Moreover, the
 ballistic update always prevails simple Euler although
 both are of first order.

In the next section, we 
specify the integration schemes in the context of car-following models. 
In Sect.~\ref{sec:sim}, we describe the simulation tests, define
the numerical complexity as a measure for the computational burden,
and the discretization error in terms of a vector norm on
the deviations of the trajectories. In Sect.~\ref{sec:results}, we
present the simulations and results. Finally. Sect.~\ref{sec:concl}
gives a discussion and an outlook.

\section{\label{sec:math}Integration Schemes for Car-Following Models
  and their mathematical properties}
 
\subsection{Mathematical Formulation}

We start by writing the dynamics created by a 
time-continuous car-following models without explicit
reaction time as a general system of ordinary
differential equations,
\be
\label{ODE}
\abl{\vec{y}}{t}=\vec{f}(\vec{y},t).
\ee
The state vector $\vec{y}$ represents all positions and
speeds, and $\vec{f}(.)$ characterizes the specific car-following
model and possibly external data such as an externally driven leading
vehicle. 
Specifically, we consider a class of car-following models  defined by
\bea
\label{carfollowGen-x}
\abl{x_i}{t} &=& v_i, \\
\label{carfollowGen-v}
\abl{v_i}{t} &=& a\sup{mic}(s_i, v_i, v_{i-1}),
\eea
where $i=1, ..., n$ denotes the index of a fixed number $n$ of
vehicles (the
first vehicle has the lowest index), $x_i$
denotes the position of the front bumper of vehicle $i$,  $v_i$ its speed,
 and $s_i=x_{i-1}-x_i-l_{i-1}$ the bumper-to-bumper
gap where $l_{i-1}$ is the length of the leading vehicle.

A model of this class is specified by the acceleration
function $a\sup{mic}(s,v,v_l)$. The simulation scenario is specified
by the number $n$ of vehicles following each other, by the initial conditions $x_i(0)$ and
$v_i(0)$ for all vehicles $i$,  and by a boundary condition
prescribing the acceleration $a_1(v_1,t)$ of the first
vehicle $i=1$. Specifically, we consider free-flow boundary
conditions~\citep{TreiberKesting-Book},  
\be
\label{BCfree}
a_1(v_1,t)=a\sub{free}(v_1)=a\sup{mic}(\infty, v_1, v_1),
\ee
and externally prescribed leader acceleration profiles 
\be
\label{BCextern}
a_1(v_1,t)=a\sub{ext}(t).
\ee
Periodic boundary conditions,
\be
\label{BCperiod}
a_1(v_1,v_n,x_1,x_n)=a\sup{mic}(x_n+L\sub{road}-x_1-l_n, v_1, v_n)
\ee
would be possible as well, as would be external influences such as
traffic lights. However, we do not allow open boundary
conditions (sources and/or sinks) since the ensuing time dependent
vehicle number $n$ would violate the general form~\refkl{ODE}. Nevertheless, we
do not expect that open boundary conditions will influence our
results in any significant way.

In order to cast the model equations~\refkl{carfollowGen-x}
and~\refkl{carfollowGen-v} into the general form~\refkl{ODE}, we
define the state vector as 
\be
\label{statevector}
\vec{y}=\myVector{\vec{x}\\ \vec{v}}
=\left(x_1, .., x_n,v_1, ..., v_n\right)\tr.
\ee
Then, the  right-hand side of~\eqref{ODE} becomes
\be
\label{rhs}
\vec{f}(\vec{y},t)
=\myVector{\vec{v}\\ \vec{a}(\vec{x}, \vec{v},t)}
=\myVector{v_1\\ \vdots \\  v_n\\ a_1(v_1,t)\\
  a\sup{mic}(x_1-x_2-l_1, v_2, v_1)\\ \vdots\\
  a\sup{mic}(x_{n-1}-x_n-l_{n-1}, v_n, v_{n-1})} \,.
\ee
Notice that the external boundary
condition~\refkl{BCextern} makes the right-hand side non-autonomous
(the independent variable $t$ appears explicitly) while, for free or periodic
boundary conditions, the ODE is autonomous.

\subsection{\label{sec:converg}Convergence and consistency order}

The quality of explicit numerical integration method with respect to
discretization errors is generally characterized
by its consistency order $p$. A method has a \textit{local} consistency order
$p\sub{loc}>0$ within a region $ R$ of state variables $\vec{y}$ and
times $t$ if it converges to the true solution $\vec{y}(t)$ and if, for
all integration time steps $h>0$ and all $\{\vec{y},t\} \in  R$, the inequality    
\be
\label{defplocal}
\epsilon\sub{loc} \equiv
\frac{\parallel\vec{y}\sub{num}(t+h)-\vec{y}(t+h)\parallel}{h} < A
h^{-p\sub{loc}} 
\ee
is satisfied. Here, $\epsilon\sub{loc}$ is the local truncation error,
$\vec{y}\sub{num}(t+h)$ is the numerical approximation for the true initial
condition $\vec{y}(t)$, $\parallel . \parallel$ is some vector norm (see
Sect.~\ref{sec:errnorm}), $A$ is a positive prefactor, 
and the consistency order is the highest positive value of $p\sub{loc}$ for which this
inequality is satisfied. The denominator $h$ ensures that
  the decreasing numerical errors are not a trivial consequence of
  decreasing time steps $h$.

More relevant for simulations of car-following
models, however, is the \textit{global} consistency order indicating
how the cumulated discretization errors of a complete simulation run decrease
with decreasing $h$. If the simulation starts at $t=0$ with given
initial conditions $\vec{y}(0)=\vec{y}_0$ and ends at $t=T$, one
may define the global consistency order by the highest value of $p\sub{glob}$
for which
 \be
\label{defpglobal}
\epsilon\sub{glob} \equiv \parallel \vec{y}\sub{num}-\vec{y}\parallel < A h^{-p\sub{glob}}
\ee
is satisfied for a finite prefactor $A$ and all $h>0$. Here, $\parallel . \parallel$ is
a vector norm on \emph{all} the components of $\vec{y}$ for all time steps,
see~\refkl{errorTraj} or~\refkl{errorGlob} below. 

If the right-hand side $\vec{f}(\vec{y},t)$ is
Lipshitz continuous in the whole region $R$ covered by the
trajectories, one can show~\citep{NumMathe-Springer} that
(i) the true solution exists, (ii) it is
unique, and (iii) converging numerical methods have a unique
consistency order
$
p\sub{loc}=p\sub{glob}=p.
$
The function $\vec{f}(\vec{y},t)$ is Lipshitz continuous if it is
differentiable with
respect to the components of $\vec{y}$ nearly everywhere and if the
gradients are bounded. We notice that, for car-following models, these
are  nontrivial requirements. For 
example, they are not satisfied for any car-following model with a
discontinuous acceleration function. In contrast, a non-differentiability at
certain points, typically introduced by min- or max conditions, is
allowed. Even for a perfectly smooth acceleration function as that of
the IDM, the Lipshitz condition is violated for gaps $s\to 0$ (crashes)
or diverging speeds or speed differences. While the latter can be
excluded by the general dynamics of the trajectories, 
the former can only be verified a posteriori.
In the following, we will base our investigations on the global
discretization error (truncation error) $\epsilon\sub{glob}$ as
defined in~\eqref{defpglobal}.  

Finally, we note that the prefactor $A$ indicating the upper bound varies wildly with
the method and the problem at hand. Generally, $A$ increases
drastically with the order of the method, if the situation includes
abrupt changes of the state, e.g., stop-and go traffic. These
variations are of a high  practical relevance  since they imply that a
higher consistency order not necessarily leads to a higher accuracy as
we will show in Sect.~\ref{sec:results}.

\subsection{The investigated integration methods}

We investigate (i) the simple Euler's method, (ii) the trapezoidal
rule (Heun's method), (iii) the standard fourth-order 
Runge-Kutta method (RK4), and (iv) the ballistic
update.  For reference, the methods (i) - (iii) for
integrating~\eqref{ODE} are  as 
follows:
\bea
\label{euler}  \text{Euler:} \quad  &&
    \vec{k_1} = \vec{f}(\vec{y},t), \nonumber \\
 && \vec{y}(t+h) = \vec{y} + h \vec{k_1}, \\ 
\label{trapez} \text{trapezoidal:} \quad  &&
    \vec{k_1} = \vec{f}(\vec{y},t), \quad
    \vec{k_2} = \vec{f}(\vec{y}+h\vec{k_1}, t+h), \nonumber \\
 && \vec{y}(t+h) = \vec{y} + \frac{h}{2} \left(\vec{k_1}+\vec{k_2}\right),\\
\label{RK4} \text{RK4:}  \quad && 
    \vec{k_1} = \vec{f}\left(\vec{y},t\right), \quad
    \vec{k_2} = \vec{f}\left(\vec{y}+\frac{h}{2}\vec{k_1}, t+\frac{h}{2} \right),
        \nonumber \\
 && \vec{k_3} = \vec{f}\left(\vec{y}+\frac{h}{2}\vec{k_2}, t+\frac{h}{2} \right), \quad
    \vec{k_4} = \vec{f}\left(\vec{y}        + h \vec{k_3}, t+h \right), 
       \nonumber \\
&& \vec{y}(t+h) = \vec{y} + \frac{h}{6}
       \left(\vec{k_1}+2\vec{k_2}+2\vec{k_3}+\vec{k_4}\right).
\eea
The ballistic method is only defined for the special case that the
ODE~\eqref{ODE} represents dynamic acceleration equations for one or
several particles which, of course, includes time-continuous
car-following models. The ballistic method assumes constant
accelerations during one time step which will be taken as that 
at the beginning of this step:
\be
\label{ballistic}
\vec{y}(t+h) =\myVector{\vec{x}(t+h)\\ \vec{v}(t+h)}
 = \myVector{\vec{x}\\ \vec{v}}
 + h \myVector{\vec{v} \\ \vec{a}(\vec{x},\vec{v}) }
 + \frac{1}{2} h^2 \myVector{\vec{a}(\vec{x},\vec{v}) \\ \vec{0}}\,.
\ee
This can be interpreted as a mixed first-order, second-order
update consisting of an Euler update for the speeds, and a trapezoidal
update for the positions. While the resulting order $p=1$ is that of
Euler's method,
it turns out that the prefactor $A$ is significantly lower in nearly
all situations. As for the Euler
update, the acceleration function $\vec{a}(\vec{x},\vec{v})$
needs only
to be calculated once per update step while two and four calculations
are necessary for the trapezoidal and RK4 updates, respectively.
Since calculating the acceleration function represents the essential
part of the numerical complexity, this gives a hint at the numerical
efficiency of the ballistic update.

\subsection{\label{sec:implNote}Special Treatment for Stopping Vehicles}

To be fully effective, the general numerical methods assume smoothness
conditions on the model 
and the data that are rarely given when
simulating car-following models. Regarding a common source of such
discontinuities, vehicle stopping, we can nevertheless improve all methods in a
systematic way by overriding the canonical formulation for such a situation.  
Due to the finite  update times, all 
update formulas~\eqref{euler} - \eqref{ballistic}
will lead to negative speeds whenever a time step includes
the stopping of vehicles. In this case, it would be
better to estimate the stopping position directly. Specifically, we
have applied, for all methods, following heuristics:
\bi
\item The special treatment is activated if, for a vehicle $i$, the speed of a predictor
  or the final step of an integration 
  scheme is negative indicating that this vehicle has stopped at some
  time instant of this time step. It also implies that the  acceleration $ a_i\sup{mic}(t)$  calculated at the
begin of this time step is negative. 
\item In case of activation, we override the originally
  calculated position by the ballistic heuristics
\be
\label{newPosCorr}
x_i(t+\tilde{h})=x_i(t)-\frac{v_i^2(t)}{2 a_i\sup{mic}(t)} \quad \text{if}
\quad v_i(t) + \tilde{h} a_i\sup{mic}(t)<0.
\ee
Here, $\tilde{h}$ is either
$h$ or $h/2$ (for the second and third predictor of RK4). 
\item Additionally, in case of activation, we reset the speed to zero.
\ei
This provision for stopping vehicles fits naturally to the
ballistic approach but improves the other methods as well. For the
higher-order methods, it increases the
maximum effective consistency order from $p=1$ to~2
 and decreases the absolute 
error, at least, if stopped vehicles are the only reason for non-smooth trajectories.

\section{\label{sec:sim}Methodology for Assessing the Integration Schemes}

In order to assess the numerical
schemes, we need to specify the numerical complexity, the global
discretization error, and the reference solution against which to
calculate the global error. Since the true solution cannot be
obtained for any nontrivial scenario (otherwise, there would be no
need for simulation), setting up the reference (or, more precisely, an
estimator for the reference with controlled errors),  is a nontrivial task.

\subsection{\label{sec:numcomplex}Numerical complexity}

We define numerical complexity as the computation time to simulate a
single vehicle  on a single lane over a given simulated time interval, e.g., \unit[1]{s}.
The inverse of this 
quantity indicates the number of vehicles that can be simulated
in real time. Since nearly all of the computational burden consists in
evaluating the model's acceleration function and a model of nominal
consistency order $p$ needs $p$ calculations per time step, 
the numerical complexity is
essentially proportional to the quantity 
\be
\label{C}
C=\frac{p}{h}
\ee
denoting the number of evaluations of the acceleration function per
vehicle and simulated unit time. The nominal consistency order is
$p=1$ for the Euler and ballistic updates, $p=2$ for the trapezoidal rule, and
$p=4$ for RK4.

\subsection{\label{sec:refSol}Reference Solution}

While a unique exact global
solution exists for our simulation scenarios
(cf. Sec.~\ref{sec:converg}), it cannot be calculated analytically for
any but the most trivial situations. We therefore calculate a
``reference solution'' $\vec{y}\sub{ref}(t)$ against which to compare the integration
schemes by the RK4 scheme using a time step
$h\sub{ref}=\unit[10^{-4}]{s}$ which is smaller than 
the smallest time step of the actual investigations by a factor of~200. To test
the validity of this reference, we repeat the calculation with
$h=2h\sub{ref}$ resulting in $\vec{y}\sub{cmp}(t)$ and calculate the global error between these two
solutions. Assuming that the \textit{actual} global consistency order of RK4
for a given scenario is
$p\sub{act}\ge 1$ (which can only be confirmed \textit{a posteriori})
this provides an upper bound for the global error between the
reference and the unknown true solution~\citep{NumMathe-Springer}:
\be
\label{ref-error}
\parallel \vec{y}\sub{ref}-\vec{y}\parallel \le 
\parallel \vec{y}\sub{cmp}-\vec{y}\sub{ref}\parallel
\ee
\begin{minipage}{\textwidth}
This controlled error of the reference solution ensures that all
global discretization errors calculated in the following can be
determined with uncertainties (i.e.,
second-order errors) of less than \unit[1]{\%}.
\end{minipage}

\subsection{\label{sec:errnorm}Global Discretization Error}

We have tested several global error norms, among them the 1-norms and
2-norms of the time series of location, speed, acceleration,
and gap for one or more trajectories, and combinations
thereof. Examples include the 1-norm of the speed trajectory of the
$i\sup{th}$ follower,
\be
\label{errorTraj}
\epsilon_i = \parallel v_i\sup{num}-v_i\sup{ref}\parallel
=\frac{1}{m}\sum\limits_{j=1}^m |v_{i}\sup{num}(jh)-v_{i}\sup{ref}(jh)|,
\ee
where $v_i\sup{num}(jh)$ is the speed of the
$i\sup{th}$ vehicle at time $t=jh$ (after
the $j\sup{th}$ time step), and $v_i\sup{ref}(jh)$ is
the  corresponding speed of the reference solution.

Another example is the 1-norm of the speeds of all  trajectories,
\be
\label{errorGlob}
\epsilon = \parallel \vec{v}\sup{num}-\vec{v}\parallel
=\frac{1}{nm}\sum\limits_{i=1}^n\sum\limits_{j=1}^m
 |v_{i}\sup{num}(jh)-v_{i}\sup{ref}(jh)|,
\ee
where $\vec{v}(t)=(v_1(t), ..., v_n(t))\tr$.

Since we always obtained
similar results, regardless of the quantity (speed or gap) or the chosen
vehicle trajectory (or including all trajectories),
 we will, henceforth, only consider $\epsilon_{10}$,
i.e., the discretization errors of the speeds of the $10\sup{th}$ vehicle.

\section{\label{sec:results}Results}

As main test cases for the simulations, we have used the city start-stop
scenario as described in Chapter~11 of~\cite{TreiberKesting-Book}:
Initially, a queue of 20~identically parameterized 
cars is waiting behind a red traffic light. At $t=0$, the traffic
light turns green and the queue of cars 
starts moving until a red traffic light $\unit[670]{m}$ ahead stops
the vehicle platoon again. Most tests are performed with the
IDM or variants thereof 
parameterized according to either the second or the third column of
Table~\ref{tab:param}.

In the theoretical  ``best case'', both the acceleration function of
the model and, if applicable, external data (such as externally
prescribed leading trajectories) are smooth, i.e., infinitely
often differentiable. This will produce smooth trajectories for which  
 all  mathematical conditions for the theoretical
consistency order are satisfied. We will start with
this ideal case before we simulate less ideal (and more
realistic) scenarios by progressively reducing the degree of
smoothness.

\begin{table}
\begin{tabular}{lll} \hline
Parameter & Standard set & ``creep-to-stop'' set \\ \hline
desired speed $v_0$ & \unit[15]{m/s} &  \unit[15]{m/s} \\[-0.5em]
time gap parameter $T$ & \unit[1]{s} & \unit[1]{s} \\[-0.5em]
minimum gap $s_0$ & \unit[2]{m} & \unit[1]{m} \\ \hline
maximum acceleration $a$ & $\unit[1]{m/s^2}$ & $\unit[2]{m/s^2}$ \\[-0.5em]
comfortable deceleration $b$ & $\unit[1.5]{m/s^2}$ & $\unit[1.5]{m/s^2}$ \\
\end{tabular}
\caption{\label{tab:param}IDM parameters of the tests.} 
\end{table}

\begin{figure}
\fig{0.7\textwidth}{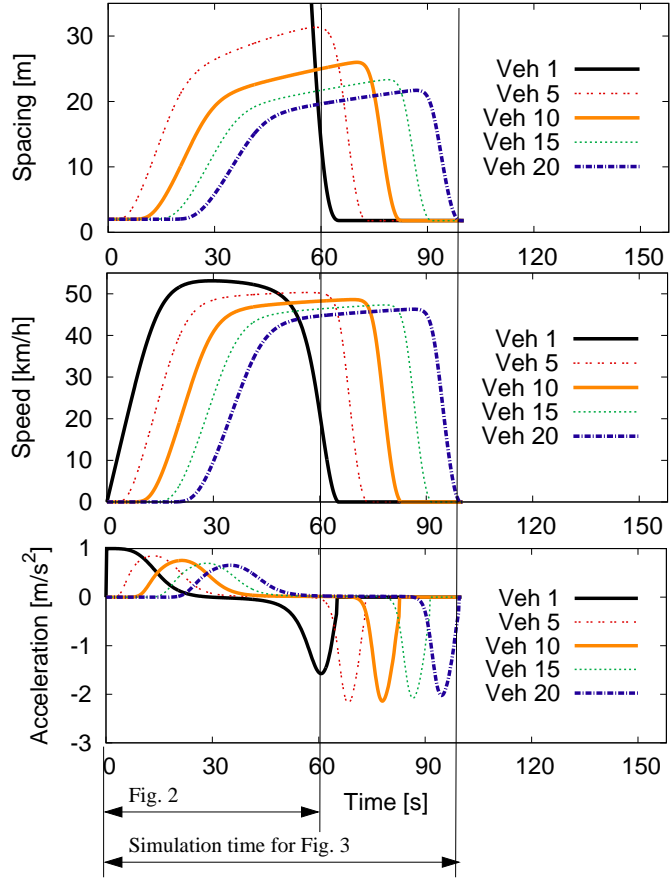}
\caption{\label{fig:scenCity}Trajectories of the start-stop
  scenario:  single-lane city traffic
  between two signalized intersections as simulated with the IDM with
  standard parameters.}
\end{figure}

\subsection{\label{sec:smooth}Smooth Trajectories}
Figure~\ref{fig:scenCity} shows  some of the resulting trajectories
when parameterizing the IDM according to the second column of
Table~\ref{tab:param}.
  When restricting the 
simulation time to \unit[60]{s}, no vehicle has stopped yet at the end of the
simulation time. Since, for nonzero speeds, the IDM
acceleration function and the resulting trajectories are
 smooth, this means that the conditions for
the maximum theoretical consistency orders are satisfied, i.e.,
 $p=1$ for Euler's and the
ballistic methods, $p=2$ for the
trapezoidal rule, and $p=4$ for RK4 method.

\begin{figure}
\fig{0.7\textwidth}{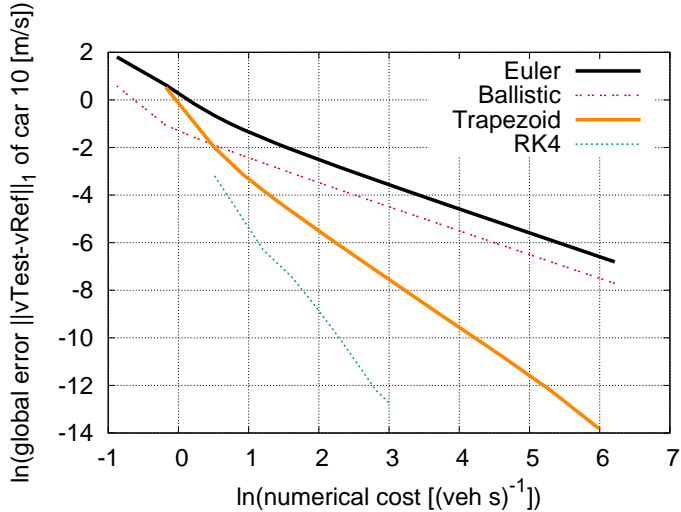}
\caption{\label{fig:numerics_cost1}Error norm of the speeds as a
  function of the numerical complexity for four numerical integration
  methods. The simulation is that of Fig.~\ref{fig:scenCity} for the
  simulation interval $[\unit[0]{s}, \unit[60]{s}]$.
}
\end{figure}

Figure~\ref{fig:numerics_cost1} shows the numerical accuracy of the
$\unit[60]{s}$-simulation in form of a log-log
plot  of the discretization
errors as a function of the numerical costs for all four integration
methods presented in Sec.~\ref{sec:math}. As error measure, we used
the 1-norm $\epsilon_{10}$ of the 
speed differences  of the $10\sup{th}$ follower. 
Specifically, we have assumed 16 different update
times ranging from $h=\unit[0.002]{s}$ to
$h=\unit[2.4]{s}$. The reference trajectory was obtained by applying
the RK4 method with an update time step
$h=\unit[10^{-4}]{s}$. We
recorded the result every \unit[2.4]{s} which is the lowest common multiple of
all update intervals $h$ used for the test simulations. This is
necessary to separate the discretization errors to be analyzed from
errors when interpolating data points and provides a common data
basis for all test simulations. As measure for the numerical cost $C$,
we defined the number of calculations of the acceleration function
per simulated vehicle and simulated second according to
Eq.~\refkl{C}. 

We observe that the
theoretical consistency orders are realized in the actual
simulation, i.e., the asymptotic negative slopes $p\sub{sim}$ of the log-log plot are
approximatively equal to $p=1$ for the Euler and ballistic
schemes, $p=2$ for the trapezoidal scheme, and $p=4$ for RK4.
Moreover, for  $h\le
\unit[0.5]{s}$ (which corresponds to $C \ge 2p$ and includes the
practically used intervals), the order of performance
of the methods (from best to worst) is RK4, trapezoidal, ballistic, and
Euler. Notice that the ballistic scheme is always superior to Euler's
scheme (only about \unit[30]{\%} of the error of the latter) 
although both have the same consistency order $p=1$.

\begin{figure}
\fig{\textwidth}{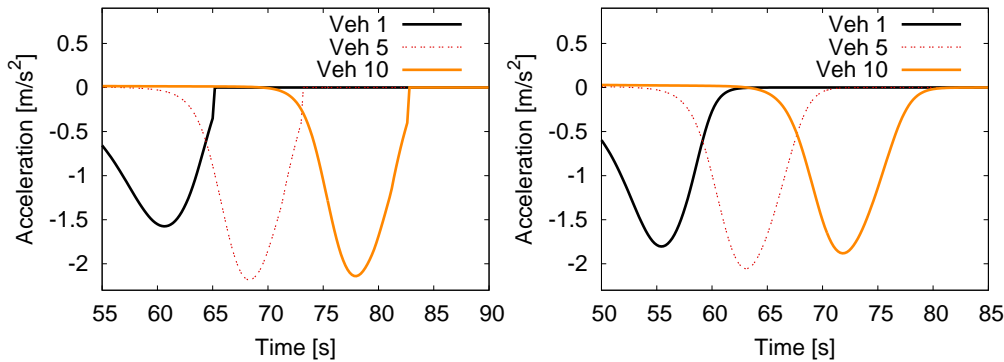}
\caption{\label{fig:creep-noCreep}Detail of the acceleration profile
  during the stopping phase. Left: distinct stop (parameterization by the 
second column of Table~\ref{tab:param}); right: creeping stop 
(third column)}
\end{figure}

\begin{figure}
\fig{0.7\textwidth}{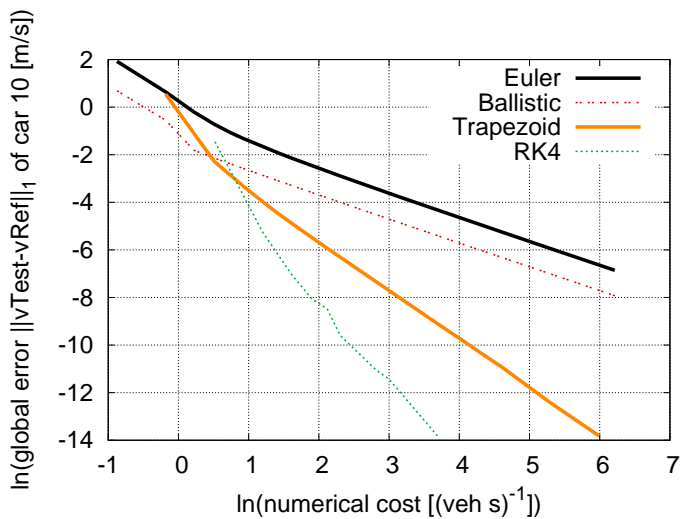}
\caption{\label{fig:numerics_cost2}As Fig.~\ref{fig:numerics_cost1}
  but for a total simulation interval of \unit[100]{s}.
}
\end{figure}

\subsection{\label{sec:stop}Discontinuous Accelerations Caused by Stopped Vehicles}

We simulate the same city start-stop scenario as above with
the same parameter set, but now for a  simulation time of 
  $t\sub{max}=\unit[100]{s}$ instead of \unit[60]{s}.
 As shown in Fig.~\ref{fig:creep-noCreep}, some vehicles have
 stopped with a  discontinuous
  acceleration profile after this prolonged time. For the higher-order
  methods, this means that the
  mathematical smoothness conditions for the theoretical consistency
  order are no longer satisfied.

Does this have implications for the
  actual accuracy? Figure~\ref{fig:numerics_cost2} shows that the
  consistency orders of the Euler, ballistic, and trapezoidal schemes
  essentially retain their theoretical values of $p=1$, 1, and 2,
  respectively. Furthermore, the prefactors $A$ of
  Eq.~\refkl{defpglobal} determining the absolute size of the discretization errors
   are essentially unchanged as well. 
In contrast, the errors of the RK4 method significantly increase by a
factor of about five. Nevertheless, RK4 remains the best method for
practical update intervals. Remarkably, its simulated consistency
order $p\approx 3.5$ is only marginally below its theoretical
value of~4. At first sight, this is unexpected since, for mathematical
reasons, the consistency 
order for a discontinuous acceleration profile should not exceed
$p=1$, regardless of the method.

There are two reasons to explain these findings. Firstly, the
discontinuity concerns just a single point, so the error contribution
of lower consistency order is small and the \textit{asymptotic} slope
may not yet have been reached in the log-log plots. 
Secondly, we have taken special provisions to increase the
accuracy of the stopping situation:  Whenever a predictor or the final
value of an integration step yields a negative speed, we override the
normal algorithm by setting the speed to zero and the position to the
ballistically  estimated 
stopping position which is calculated assuming a constant deceleration
defined by the right-hand side of~\refkl{rhs} for this step.
This special-purpose procedure, which is described in
Sec.~\ref{sec:implNote} below in more detail, increases the upper bound of the expected
consistency order to~2 in all simulations that include stops  but
have  smooth acceleration functions and data, otherwise.

\begin{figure}
\fig{0.7\textwidth}{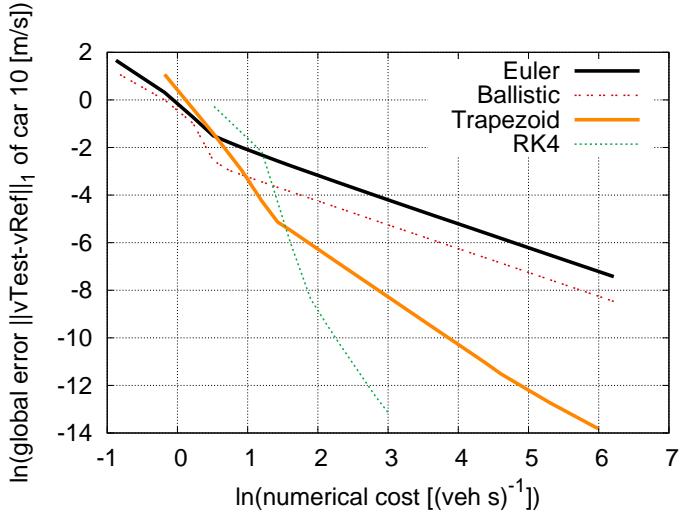}
\caption{\label{fig:numerics_creep}Discretization errors of the
  scenario of Fig.~\ref{fig:numerics_cost2} when
  re-parameterizing the IDM by the $3\sup{rd}$ column of Table~\ref{tab:param}
    resulting in a creeping halt.
}
\end{figure}

\begin{figure}
\fig{0.7\textwidth}{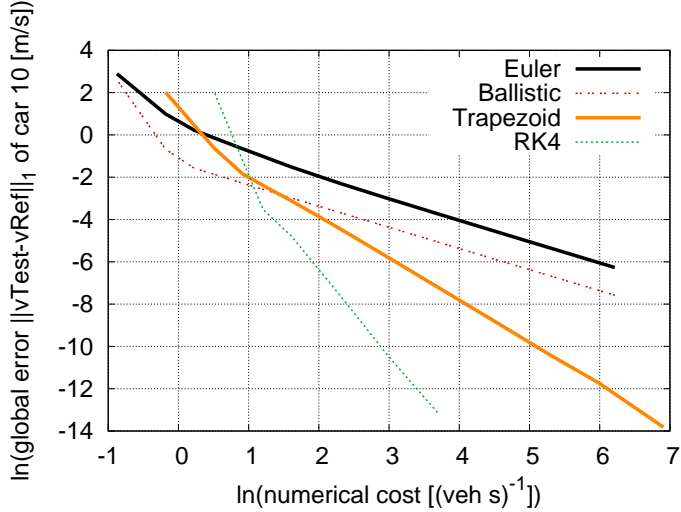}
\caption{\label{fig:numerics_OVM}Discretization errors when simulating the scenario of
  Fig.~\ref{fig:numerics_cost2} with the optimal-velocity model
  (creeping halt).
}
\end{figure}

Depending on the parameterization, it is
possible that the simulated vehicles do not stop at a precisely
defined time (as above) but ``creep to a halt'' retaining a smooth acceleration
profile at all times. An example parameterization for this behaviour
is given by the third column  of
  Table~\ref{tab:param} resulting in the trajectories of 
Fig.~\ref{fig:creep-noCreep} right. Then, the
mathematical conditions for the full theoretical consistency order are
satisfied again. In fact, the log-log plot of the discretization error
vs. numerical cost (Fig.~\ref{fig:numerics_creep}) shows little
changes with respect to the first simulation without stops.
Particularly, the slopes are consistent with the theoretical
expectation again. However, the
RK4 method produced significantly larger errors compared to the
\unit[60]{s} simulation when simulating with comparatively large update time
intervals.  We also simulated this scenario with the OVM
of~\cite{Bando1} with typical parameters resulting in a creeping halt
as well. Again, we found the expected theoretical consistency orders but
a higher prefactor, i.e., generally higher errors for all methods and
all discretization time steps.

\subsection{\label{sec:kinkAcc}Non-Smooth Acceleration Functions}

Discontinuous acceleration profiles do not only result when vehicles
stop  but may also be generated by the model itself if its
acceleration function $a\sub{mic}(s,v,v_l)$ has regions in state space
with kinks
(non-differentiable points) or discontinuities, at least when the
dynamics reaches these regions. We tested both cases by simulating the standard
start-stop scenario ($t\sub{max}=\unit[100]{s}$) with two
modifications of the IDM. As an example for an acceleration function with
kinks, we simulated the ``IDM-Plus'' proposed
by~\cite{IDMplus}. Its acceleration function reads 
\begin{figure}
\includegraphics[width=0.49\textwidth]{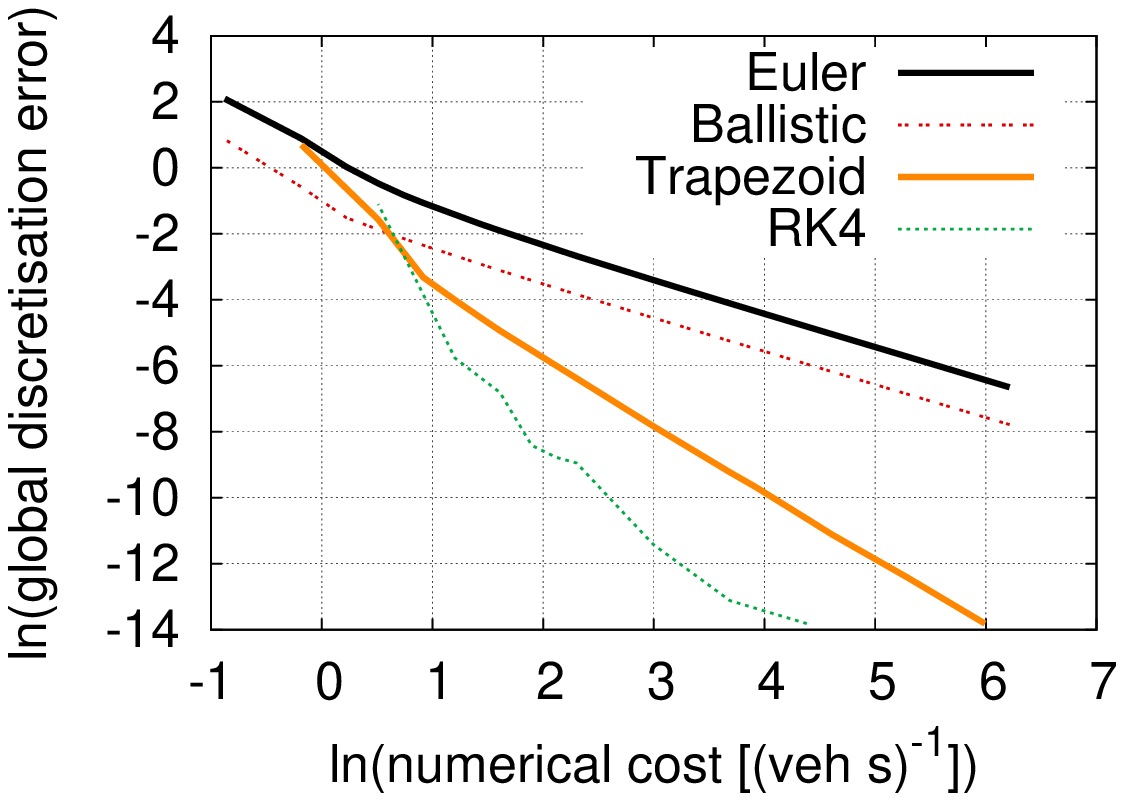}
\includegraphics[width=0.49\textwidth]{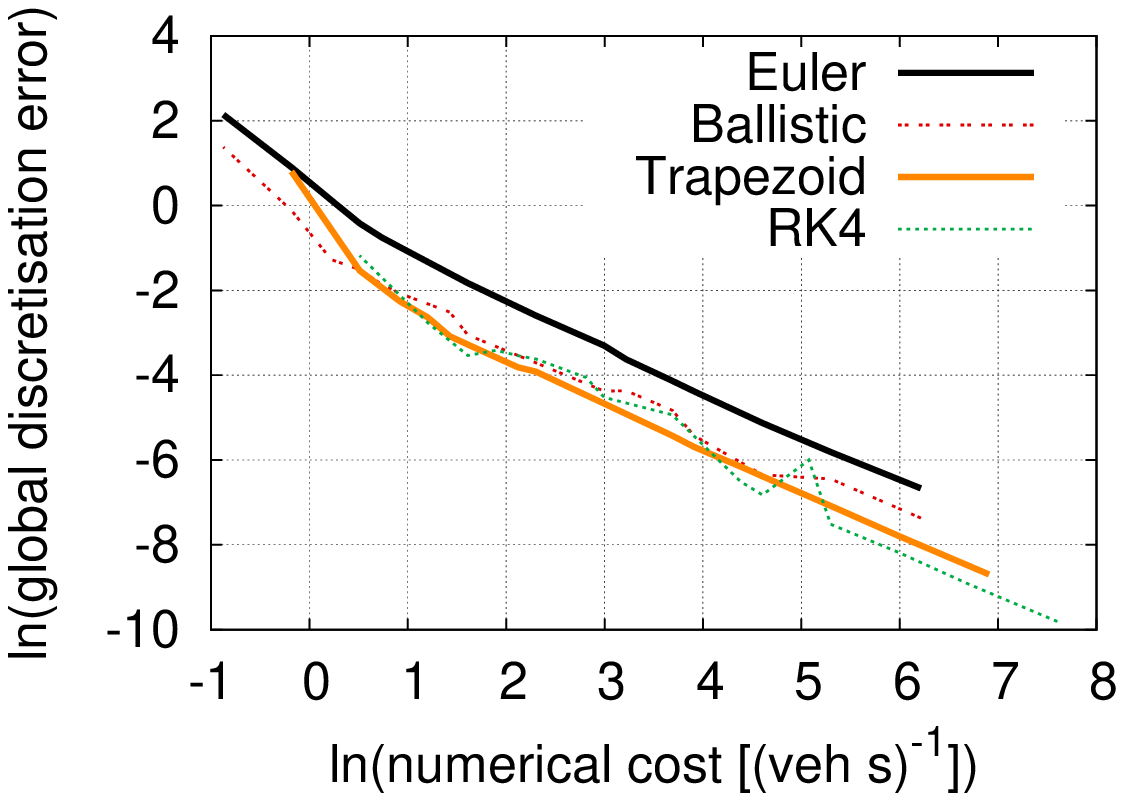}
\caption{\label{fig:numerics_IDMvariants}Discretization errors of the start-stop scenario
  ($t\sub{max}=\unit[100]{s}$) for the IDM-Plus (left) and the  modified IDM according
  to~\refkl{IDMmod} (right). All IDM variants are parameterized according to the
  second column of Table~\ref{tab:param}.
}
\end{figure}

\be
\label{IDMplus}
a\sup{mic}\sub{IDM+}(s,v,v_l)=\min\left\{a\sub{free}(v), 
a \left[1-\left(\frac{s^*(v,v_l)}{s}\right)^2\right]\right\}
\ee
with the usual IDM expressions for the free acceleration and the dynamic desired gap,
\be
\label{afree-sstar}
a\sub{free}(v)=a\left[1-\left(\frac{v}{v_0}\right)^4\right], \quad
s^*(v,v_l)=\max\left[s_0+vT+\frac{v(v-v_l)}{2\sqrt{ab}}, 0\right].
\ee
For reference, the acceleration function of the original IDM reads
\be
\label{IDM}
a\sup{mic}\sub{IDM}(s,v,v_l)=a\sub{free}(v)- a \, 
\left(\frac{s^*(v,v_l)}{s}\right)^2.
\ee
The additional ``min'' condition of the IDM-Plus\footnote{The ``min''
  function of the dynamic desired gap of all IDM variants becomes only relevant in rare
  cases.}  leads to a kink in the
acceleration profile when the remaining distance of the first vehicle
to the red traffic light (modelled as a standing virtual vehicle of length
zero) is  
approximatively $s_c=s_0+v_0T+1/2 v_0^2/\sqrt{ab}$, and also to
(smaller) kinks of the 
accelerations of the following vehicles. 

Figure~\ref{fig:numerics_IDMvariants} (left) shows the
discretization errors for the IDM-Plus. We observe that both the
error prefactors and the 
consistency orders of the Euler, ballistic, and trapezoidal methods
remains essentially unchanged while the consistency order of RK4
reduces to about two. This is expected on theoretical grounds: A kink
in the realized acceleration profile sets the upper limit of the
consistency order of any explicit method to $p\sub{max}=2$.
For practical simulation time intervals
corresponding to a numerical cost of around $\unit[10]{(veh\,
  s)^{-1}}$, however, we observe little difference with respect to
discretization errors for a smooth acceleration profile.

To obtain a model with \textit{discontinuities} in the acceleration function, we modified the
free-flow IDM acceleration function to
\be
\label{IDMmod}
a\sub{free}(v)=\twoCases{a}{v<v_0}{1-v/v_0}{v\ge v_0}. 
\ee
In this model, drivers reduce their acceleration
abruptly from $a$ to zero once reaching their desired speed. 

Figure~\ref{fig:numerics_IDMvariants} (right) shows that
acceleration discontinuities greatly increase the discretization errors of the
higher-order methods. In line with theoretical expectations, all
methods now have the consistency order $p=1$. Regarding the absolute
errors, Euler is worst while all other methods are essentially
equivalent. 

\begin{figure}
\fig{0.8\textwidth}{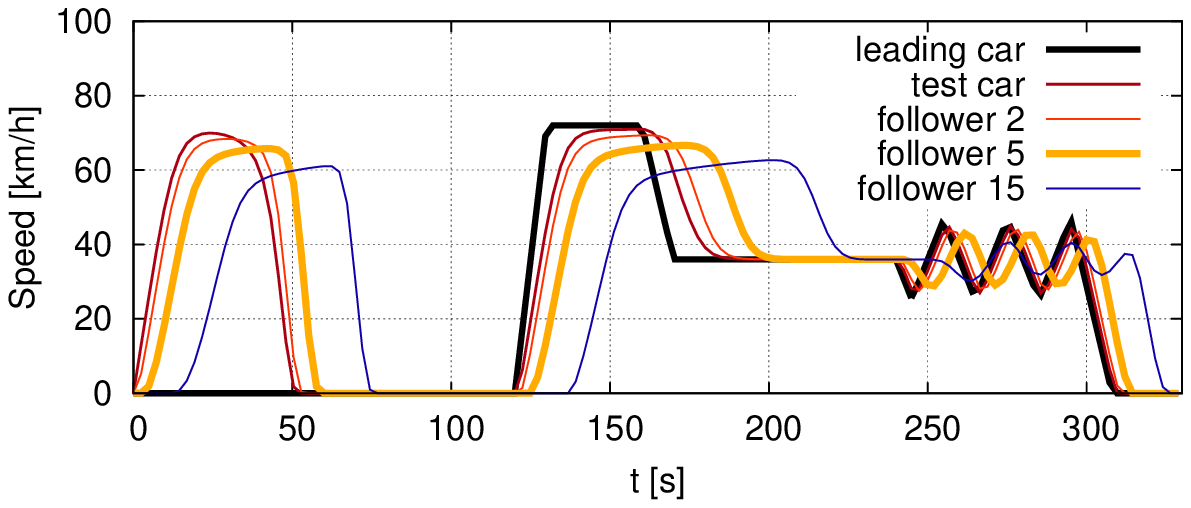}
\fig{0.7\textwidth}{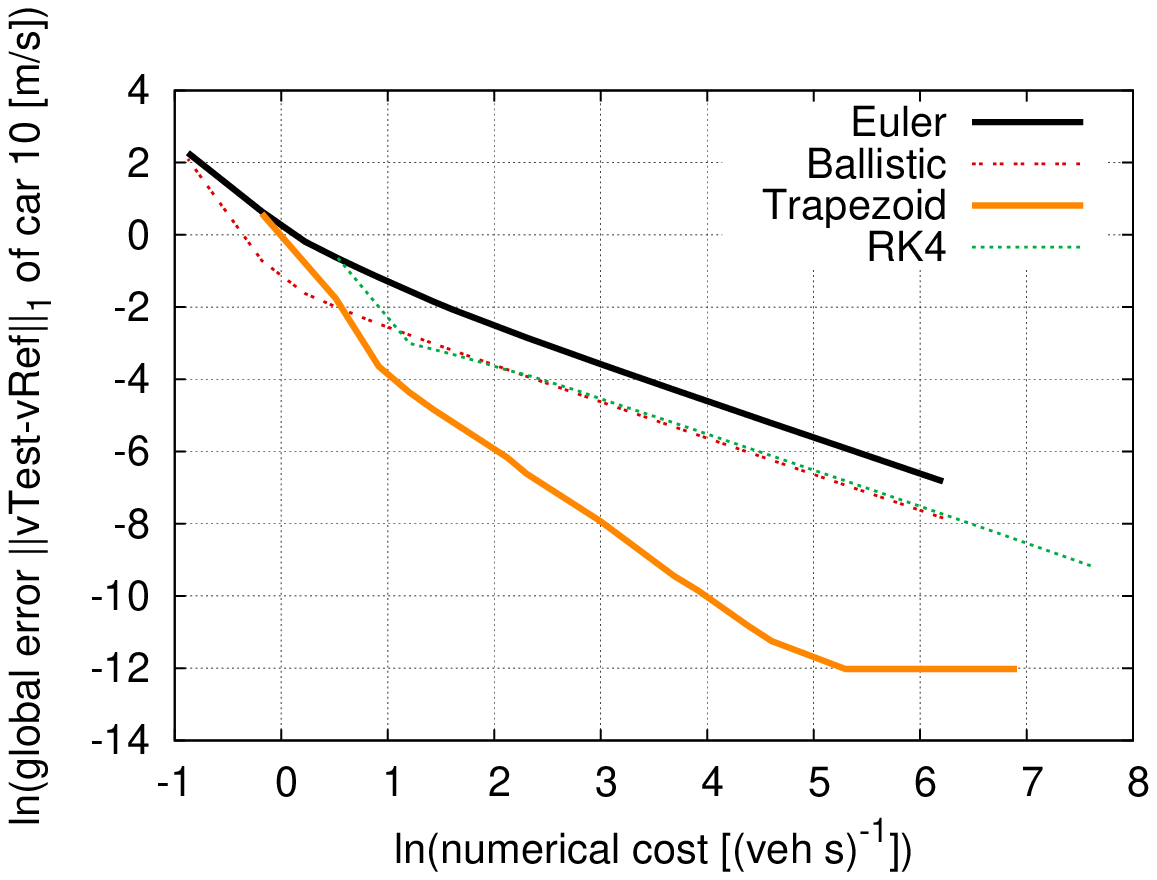}
\caption{\label{fig:fixLeader}Top: following  a leader with a fixed
  speed profile (IDM, parameters as in
  the second column of Table~\ref{tab:param}); bottom: convergence
  diagram for the tenth follower. 
}
\end{figure}

\subsection{\label{sec:discData}External Data with Discontinuous Accelerations}

Another source of discontinuities can be external system data, e.g.,
prescribed speed profiles of an external leader.
Figure~\ref{fig:fixLeader} shows a simulation where a platoon of
vehicles follows a leader with an externally given discontinuous
acceleration profile corresponding to a speed profile with
kinks. Assuming a model with a continuous acceleration function, this
leads to an acceleration profile with kinks for the first follower,
and to differentiable profiles for the further followers.

The log-log plot of the discretization errors
(Fig.~\ref{fig:fixLeader} bottom) reveals that the Euler, ballistic,
and trapezoidal methods have their expected consistency orders and
absolute errors while, surprisingly, RK4 has only consistency order
$p=1$. It seems that, at each abrupt behavioral change of the leader,
the predictors of RK4 err to such an extent that the result is
essentially that of the ballistic method.

\begin{figure}
\fig{0.8\textwidth}{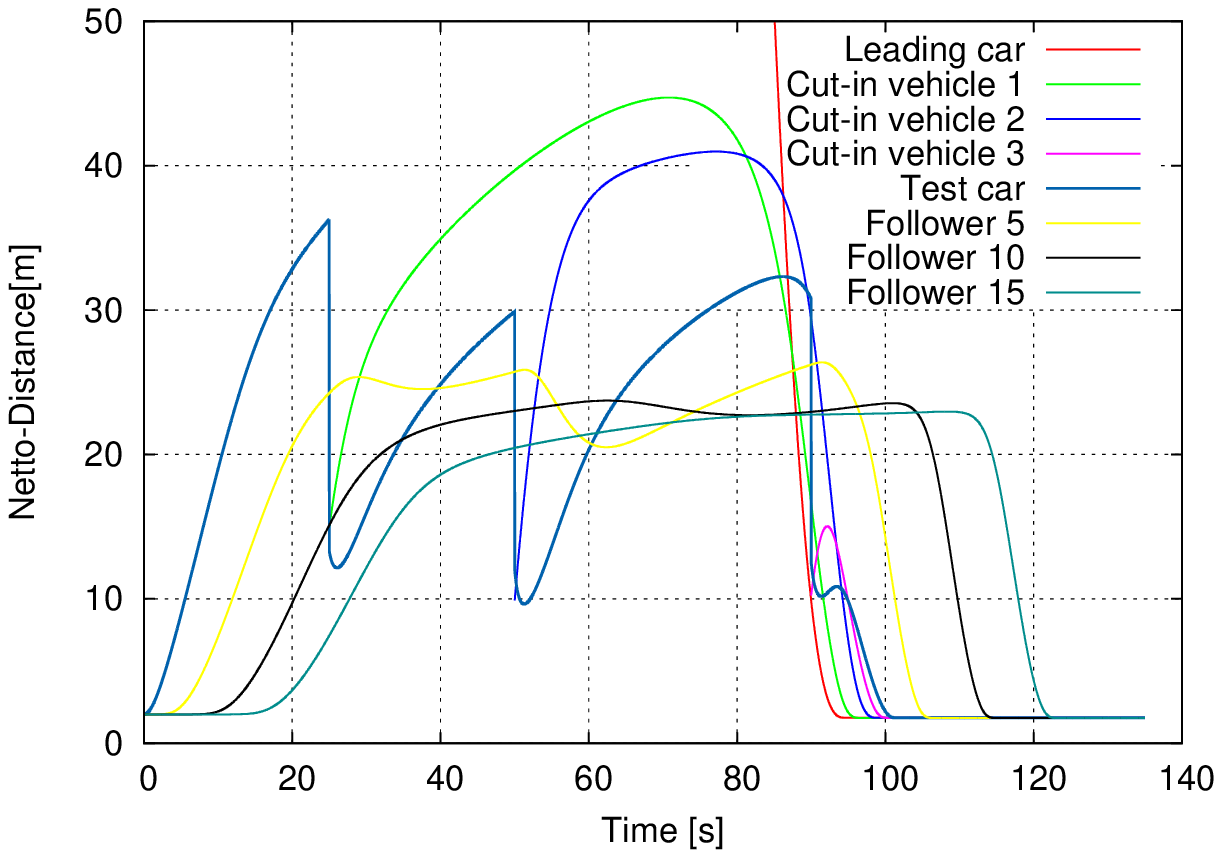}
\fig{0.8\textwidth}{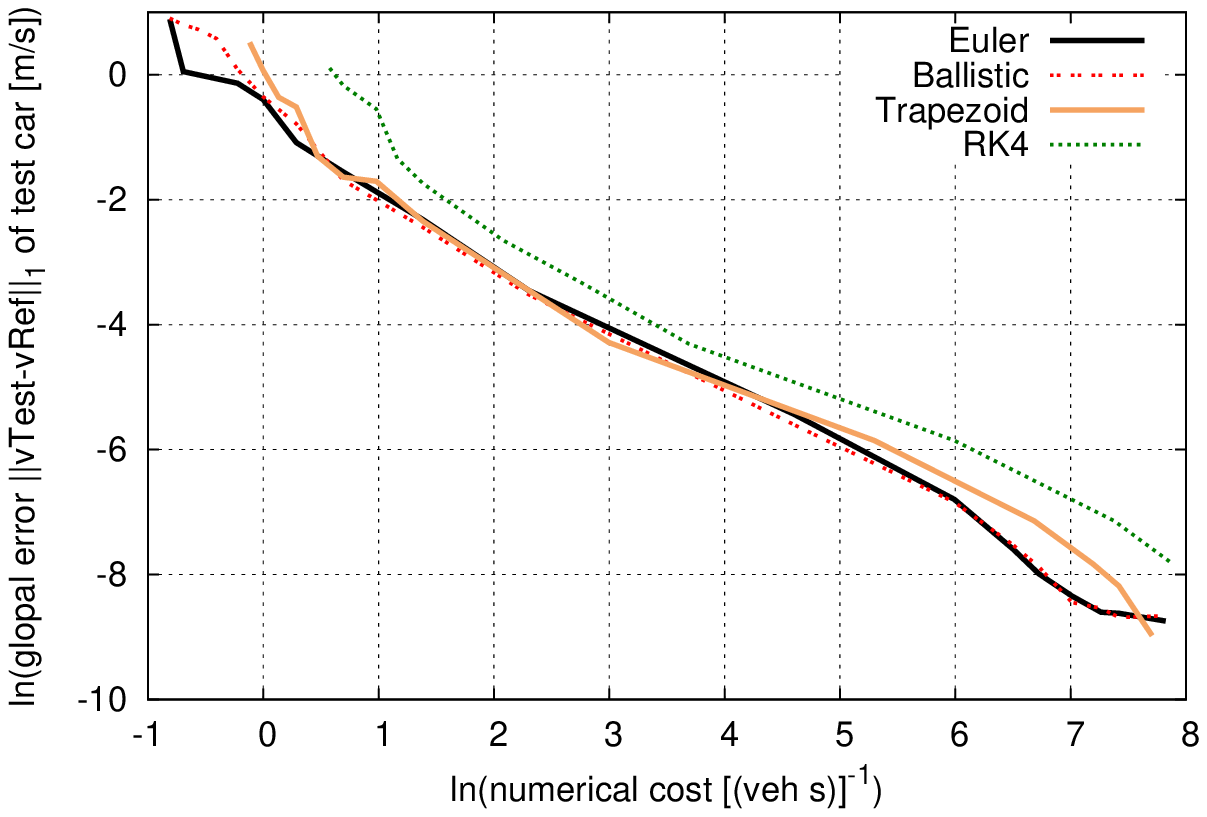}
\caption{\label{fig:cutin}Top: gap in front of the test vehicle. The
  three discontinuities represent three vehicles merging in front of
  this vehicle; bottom: convergence
  diagram for the test vehicle.
}
\end{figure}

\subsection{\label{sec:LC}Lane Changes}

The most severe source of discontinuities are active and passive lane
changes, i.e., the considered vehicle changes itself, or another changing vehicle
``cuts in'' in front of it. From the mathematical point of view, this
means that not only the accelerations of the leader are discontinuous
but the gaps and leading speeds as well. Since the two latter
quantities are exogenous variables of the model's acceleration
functions, the acceleration profile of the vehicle behind the
lane-changing vehicle on the target lane is discontinuous as well
reducing the consistency orders of all methods to $p=1$. 

This is, in fact,
what we observe: Figure~\ref{fig:cutin} displays a simulation where the test vehicle
encounters three cut-ins in front of it. All four integration methods
have the empirical consistency order
$p=1$. Remarkably and unexpectedly, the absolute value of the error
for a given numerical 
cost is largest for RK4.

\section{\label{sec:concl}Discussion and Conclusions}
In this work, we have systematically investigated the global 
discretization
error of several explicit numerical integration schemes commonly used for
simulating time-continuous car-following models. To enable an
equitable comparison between simple and higher-order methods, we have
determined the errors as a function of the numerical complexity, i.e., 
the normalized  computation time for simulating one vehicle over one
time unit: A method is better if, for the same  numerical complexity,
its global errors are lower.

Generally, when integrating ODEs or systems thereof,  the fourth-order Runge-Kutta
scheme (RK4) is the \textit{de-facto} standard and other methods are
rarely used. Why is this not the case for integrating car-following
models where  Euler's method is most widespread? This
contribution shows that, for typical traffic-related situations,
the RK4 method is, in fact,  not the best
method since the smoothness 
conditions to reach its theoretical consistency order
$p=4$ are rarely given. To investigate the effect of violating these 
conditions, we simulated several scenarios and several models,
from the ideal case to the
most severe violations of smoothness:
\benum
\item All trajectories remain smooth (sufficiently often
  differentiable) over the complete simulation time
  (Sect.~\ref{sec:smooth}), 
\item the acceleration profile of the trajectories is continuous but
  not smooth due to the model's acceleration function
  (Sect.~\ref{sec:kinkAcc} or the external data (Sect.~\ref{sec:discData}),
\item the acceleration profile is discontinuous due to stopped
  vehicles (Sect.~\ref{sec:stop}) or by discontinuities in the model's
  acceleration function which are reached by the system dynamics
  (Fig.~\ref{fig:fixLeader}), 
\item the speed and gap profiles are discontinuous as a consequence
  of lane  changes (Sect.~\ref{sec:LC}).
\eenum
We have found that RK4 and the trapezoidal scheme perform best if the
trajectories are smooth or have, at most, kinks in the
acceleration. With our special treatment of stopping vehicles
(Sec.\ref{sec:implNote}), 
this also carries 
over to stops with a discontinuous acceleration. In all other
situations, however, the consistency order of \textit{all} methods is
restricted to one and the ballistic and trapezoidal schemes are equally
performant as RK4. Moreover, when including lane changes, RK4
turned out to have the worst performance, even worse than simple Euler. 

In summary,  we recommend the ballistic and trapezoidal methods as
efficient and robust schemes for integrating car-following
models. Although of the same theoretical consistency order $p=1$ as Euler's
method, the
ballistic scheme turned out to be \textit{consistently} better than Euler's
scheme with typically only about \unit[30]{\%} of the discretization
errors compared to the latter method.



\bibliographystyle{elsart-harv}

\bibliography{database,newRefsNumerics}



\end{document}